# IMPROVING THE PERFORMANCE OF MID-T BAKED NIOBIUM CAVITIES THROUGH POST-BAKE SURFACE TREATMENT*

V. Chouhan[†,1], D. Bice[1], A. Cravatta[1], B. Guilfoyle[2], A. Murthy[1], A. Netepenko[1], T. Reid[2],
T. Ring[1], D. Smith[1], G. Wu[1]
[1]Fermi National Accelerator Laboratory, Batavia, USA
[2]Argonne National Laboratory, Lemont, USA

## Abstract

The Medium temperature (mid-T) baking of niobium superconducting radio-frequency cavities at 300–350 °C in a vacuum furnace is known to enhance the quality factor ($Q_0$). However, despite this improvement, cavities treated with this process often exhibit premature quench at relatively low accelerating fields. This limitation is suspected to arise from the formation of surface contaminants, such as niobium carbides, during the furnace bake at 350 °C for 3 h. To investigate the influence of potential surface contamination, this study applied an ultralight chemical removal to 1.3 GHz and 650 MHz single-cell cavities that had undergone medium-temperature baking. The removal of the top RF surface layer led to a notable improvement in the quench field and $Q_0$, indicating a beneficial effect of eliminating possible surface residues introduced during the bake.

## INTRODUCTION

Niobium (Nb) made superconducting radio frequency (SRF) cavities are key components in a high energy particle accelerator machine. For efficient accelerator operation, these cavities must show excellent superconducting performance, characterized by a high accelerating gradient ($E_{acc}$) and a high quality factor ($Q_0$). A high $Q_0$ is particularly desirable because it minimizes cryogenic heat load, thereby reducing operational costs.

The $Q_0$ around the medium field can be enhanced by surface modification techniques such as nitrogen doping and mid-T baking, both invented at Fermilab [1, 2]. Mid-T baking is typically performed at 300–350 °C for 3 h, either in-situ or in a vacuum furnace. In-situ baking involves actively pumping the cavity's internal volume to maintain high vacuum conditions during the bake [2], whereas the furnace baking of the cavity is performed by placing the cavity in a vacuum furnace [3, 4]. The vacuum furnace baking is preferred for large-scale production.

However, mid-T baking can result in the formation of non-superconducting niobium carbides (NbC) on the surface, which may limit cavity performance. The extent to which these surface contaminants affect RF performance remains poorly understood. To address this, the present study is focused on the removal of the top RF penetration layer from mid-T baked cavities by ultralight electropolishing (EP) and compares cavity performance before and after this removal to elucidate the role of surface contaminants.

## SAMPLE PREPARATION AND ANALYSIS

To evaluate the chemical state of the mid-T baked Nb surface and the effect of post-bake ultralight EP, two high-RRR Nb samples (Sample-1 and Sample-2) were prepared following a systematic procedure. Both samples first underwent bulk EP and ultrasonic cleaning, after which they were subjected to mid-T baking at 350 °C for 3 h in the same high-vacuum furnace used for cavity degassing and mid-T baking. Typical furnace temperature and pressure profiles during mid-T baking at 350 °C for 3 h are shown in Fig. 1.

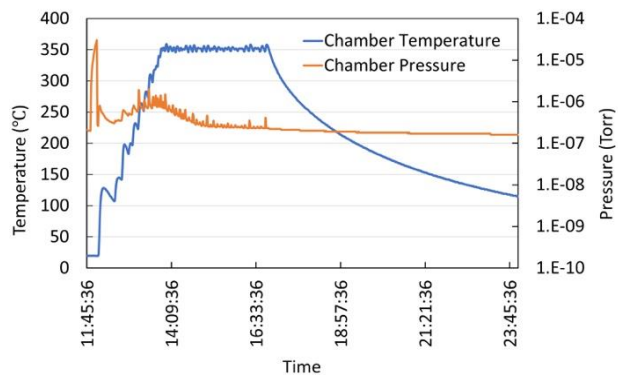

Figure 1: Typical furnace temperature and pressure profiles during mid-T baking at 350 °C for 3 h.

Following baking, Sample-2 received ultralight EP corresponding to approximately 120 nm of material removal. This step was intended to eliminate potential surface and near-surface contamination introduced during the mid-T bake. This ultralight EP was carried out using standard EP electrolyte, which is a mixture of sulfuric acid (96wt%) and hydrofluoric acid (70wt%) in a volumetric ratio of 10:1. The details on the sample EP setup have been reported in reference [5]

Both sample surfaces were analyzed using Secondary Ion Mass Spectrometry (SIMS) to obtain elemental depth profiles. SIMS results for Oxygen (O), carbon (C), and niobium carbide (NbC), each normalized to Nb intensity, are presented as a function of surface depth in Fig. 2. The Intensity of O was found to be similar for both samples. However, Sample-1 exhibited a significantly higher NbC signal than Sample-2. This comparison suggests that the top tens of nanometers of the mid-T baked surface contain NbC, which can be effectively removed by ultralight EP without altering the O content within the RF layer.

___________________
* This work has been supported by Fermi Forward Discovery Group, LLC under Contract No. 89243024CSC000002 with the U.S. Department of Energy, Office of Science, Office of High Energy Physics.
† vchouhan@fnal.gov

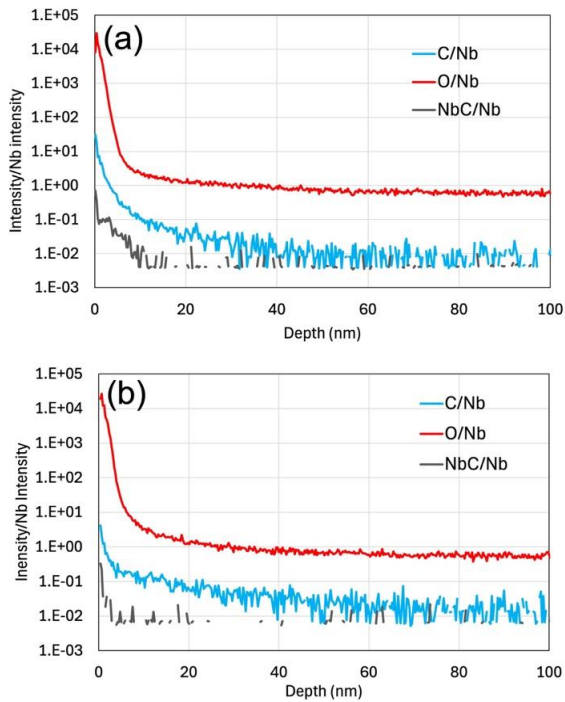

Figure 2: SIMS depth profiles of C, O, and NbC for (a) Sample-1: mid-T baked Nb and (b) Sample-2: mid-T baked Nb with ultralight EP.

## CAVITY PROCESSING

A 1.3 GHz single-cell cavity (TE1RI010) and a 650 MHz single-cell cavity (B9AS-AES-003) with β = 0.9 were prepared for baseline tests and mid-T bake studies. Both cavities underwent bulk EP, degassing at 900 °C for 3 h, light EP for 40 μm removal, and the standard low-temperature bake at 120 °C for 48 h to establish their baseline performance. Details of the EP tool used for TE1RI010 are provided in reference [6]. The baseline processing steps for B9AS-AES-003, including used optimized EP conditions and cathode geometry, along with its baseline performance, have been reported elsewhere [7].

After the baseline tests, both cavities were subjected to 5 μm light EP to remove the effect of the 120 °C bake prior to mid-T baking. Mid-T baking was conducted at 350 °C for 3 hours in the vacuum furnace. The cavity flanges were covered with Nb caps to mitigate cavity surface contamination from the furnace.

As the mid-T baked sample surface contained NbC, the cavity surface might also contain these contaminants and affect the performance. To evaluate their roles on the cavity performance, the top 100–300 nm thick surface layer was removed by the EP process. Such a small removal thickness was chosen because 120 nm EP was found to be enough to remove such contaminants from the surface, as seen from the sample analysis.

Ultralight EP was also performed with the standard EP electrolyte under controlled conditions, with the cavity temperature maintained below that used in standard cold EP to slow the removal rate and limit material removal to < 150 nm. EP current density and removal thickness as a function of elapsed EP time are shown for TE1RI010 in Fig. 3. Average material removal was estimated to be 108 nm for TE1RI010. B9AS-AES-003 received two ultralight EP for ~147 nm and ~118 nm to remove a total of 265 nm.

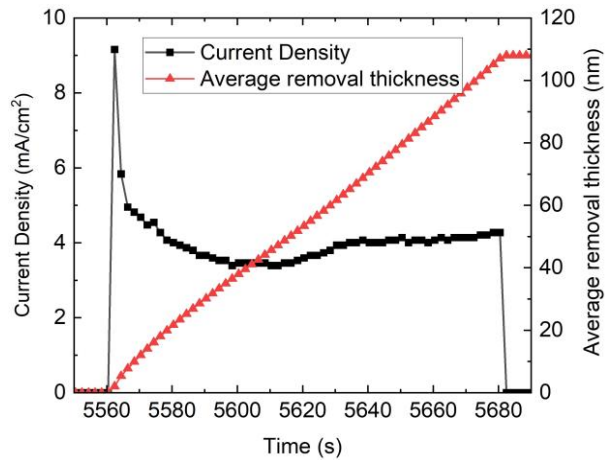

Figure 3: EP current density and average removal thickness during the EP process of TE1RI010.

## CAVITY PERFORMANCE: VERTICAL TEST RESULTS

The cavities were tested in a vertical cryostat at 2 K and lower temperatures, including ~1.5 K, for each surface condition: 120 °C baking, mid-T baking, and ultralight EP. A Helmholtz coil was used to cancel the ambient magnetic field. Additionally, fast cooldown was used for efficient flux expulsion [8].

### 1.3 GHz Single-cell Cavity

Figure 4 shows $Q_o$ of TE1RI010 measured at 2 K as a function of both peak surface magnetic field ($B_p$) and $E_{acc}$. These plots are referred to as Q-E curves in this work. The cavity after 120 °C bake reached a high gradient of 47 MV/m (~199.8 mT) before quenching. After mid-T bake, the quench field dropped significantly to 22 MV/m. Despite this low quench field, the $Q_o$ increased compared to the 120 °C baked state. However, the curve lacked the typical anti-Q slope usually observed in mid-T baked 1.3 GHz cavities.

After ultralight EP, the cavity showed enhanced $Q_o$ with a pronounced anti-Q slope. The quench field also improved, increasing from 22 MV/m to 32 MV/m. Table 1 summarizes the cavity performance after each processing step.

The results suggest that the top mid-T baked surface with a thickness of ~100 nm contained some impurities, possibly NbC, that resulted in the lower $Q_o$. These impurities might also result in premature quenching as observed after mid-T baking.

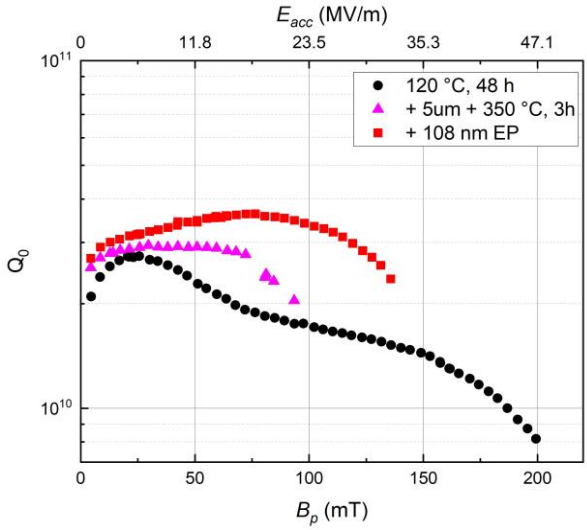

Figure 4: $Q_o$ of TE1RI010 as a function of peak surface magnetic field and $E_{acc}$. Three curves represent the performance after the 120 °C bake, mid-T bake, and ultralight EP treatment.

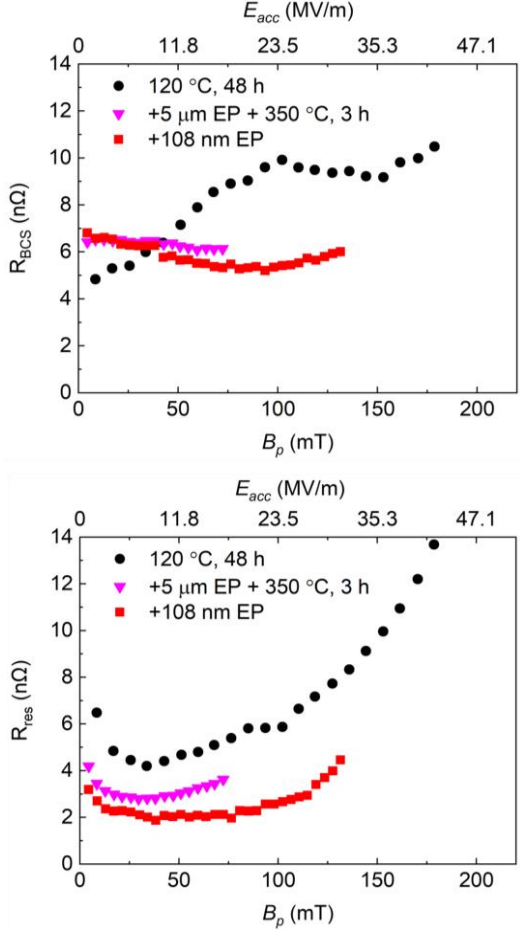

Figure 5: $R_{BCS}$ (top) and $R_{res}$ (bottom) as a function of peak surface magnetic field ($B_p$) and accelerating gradient ($E_{acc}$) for TE1RI010 cavity after the 120 °C bake, mid-T bake, and ultralight EP treatment applied following mid-T baking.

To better understand the effect of ultralight EP on the performance, the surface resistances were compared. The measured surface resistance $R_s = G/Q$ (where $G$ is the cavity's geometrical factor) was decomposed into BCS resistance $R_{BCS}$ and residual resistance $R_{res}$. The field dependence of these components for three surface conditions is shown in Fig. 5. Mid-T baking reduced both $R_{BCS}$ and $R_{res}$ relative to the 120 °C bake condition. Although $R_{res}$ remained lower than that after 120 °C bake, it began increasing from ~50 mT.

Ultralight EP resulted in further reduction in $R_{res}$, while $R_{BCS}$ remained comparable to the mid-T baked surface, with a downward trend starting near $B_p$ ~ 50 mT. Moreover, the onset of rapid increase in $R_{res}$ shifted from ~50 mT to ~100 mT. The reductions in both resistances became more pronounced at higher fields.

The comparison of resistance indicated that surface impurities mainly affected $R_{res}$, which could be minimized when the top RF layer containing impurities was removed.

Table 1: Cavity (TE1RI010) Performance at 2 K after 120 °C Bake, Mid-T Bake at 350 °C, and Post-mid-T-bake Ultralight EP

| Pre-vertical test process | $Q_o$ at 72 mT (~17 MV/m) | Quench field (MV/m) |
|---|---|---|
| 120 °C, 48 h | $1.9 \times 10^{10}$ | 47 (199.8 mT) |
| 5 μm EP + Mid-T bake | $2.8 \times 10^{10}$ | 22 (93.5 mT) |
| Mid-T bake + Ultralight EP (108 nm) | $3.6 \times 10^{10}$ | 32 (137 mT) |

*650 MHz Single-cell Cavity*

The performance of B9AS-AES-003 measured at 2 K for the three surface conditions is shown in Fig. 6. As reported earlier [7], the cavity after the 120 °C bake achieved a high gradient of 53.3 MV/m, corresponding to 198.3 mT. After mid-T baking, the quench field degraded to 33.5 MV/m (123.9 mT), and the measured $Q_o$ was lower than that obtained after 120 °C bake.

Ultralight EP for 147 nm removal significantly improved $Q_o$ in the field range of 13–27 MV/m (50–100 mT). The quench field also improved slightly to 35.3 MV/m (~131.3 mT). To further evaluate the effect of additional surface removal, the cavity was tested after the second ultralight EP (118 nm). The cavity $Q_o$ after the second ultralight EP was comparable to that achieved after the first ultralight EP, while the quench field decreased slightly to 33.1 MV/m. The similarity in $Q_o$ after both ultralight EP treatments suggests that the top < 150 nm of the surface was primarily responsible for the lower $Q_o$ observed after mid-T baking. The results are summarized in Table 2.

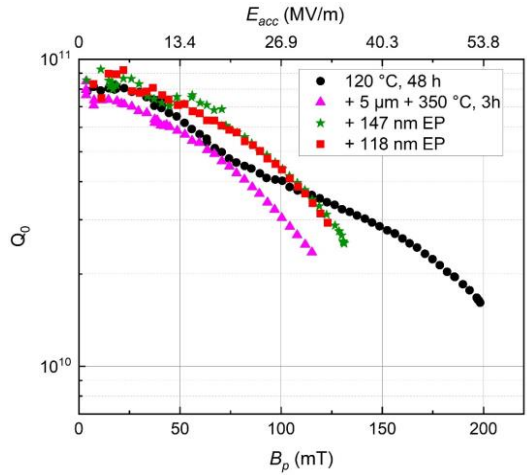

Figure 6: *Q-E* curves measured for B9AS-AES-003 cavity after 120 °C bake, mid-T bake, and two successive ultralight EP treatments applied to the mid-T baked surface.

Table 2: Cavity Performance at 2 K after 120 °C Bake, Mid-T Bake at 350 °C, and Post-mid-T-bake Ultralight EP

| Pre-vertical test process | $Q_o$ at 72 mT (~19 MV/m) | Quench field (MV/m) |
|---|---|---|
| 120 °C, 48 h | $5.0 \times 10^{10}$ | 53.3 (198.3 mT) |
| 5 µm EP + Mid-T bake | $4.7 \times 10^{10}$ | 33.5 (124.6 mT) |
| 1st Ultralight EP (147 nm) | $6.1 \times 10^{10}$ | 35.3 (131.3 mT) |
| 2nd Ultralight EP (118 nm) | $6.1 \times 10^{10}$ | 33.1 (123.1 mT) |

The calculated $R_{BCS}$ and $R_{res}$ for all three surface states are presented in Fig. 7 for comparison. For the mid-T baked surface, $R_{BCS}$ was higher compared to the 120 °C bake case at fields below 50 mT and above 75 mT, while it was slightly higher in the 50–75 mT range. The mid-T bake reduced $R_{res}$ at the fields below 75 mT and increased it at higher fields. Further study is required to understand why $R_{BCS}$ was higher after the mid-T bake.

After the two ultralight EP treatments, $R_{res}$ decreased compared to the mid-T bake condition, with the reduction becoming more significant at higher fields. $R_{BCS}$ remained similar up to ~50 mT but significantly decreased at fields above ~50 mT.

Since $R_{BCS}$ scales with $f^2$ (where *f* is the cavity's resonant frequency), the mid-T baked 650 MHz cavity shows a lower contribution from $R_{BCS}$ compared to the 1.3 GHz cavity. The unexpected lower $Q_o$ of the mid-T baked B9AS-AES-003 around the medium field ~75 mT was attributed to a higher $R_{res}$ caused by the surface impurities. As observed, ultralight EP lowered $R_{res}$ and improved $Q_o$. The effect of ultralight EP on the surface resistances for the 650 MHz cavity appeared consistent with that observed for the 1.3 GHz cavity.

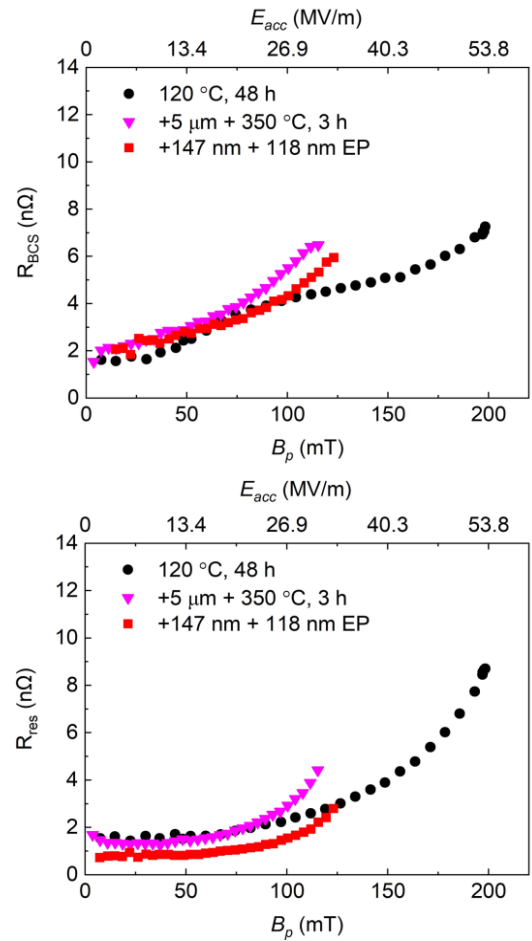

Figure 7: $R_{BCS}$ (top) and $R_{res}$ (bottom) as a function of peak surface magnetic field ($B_p$) and accelerating gradient ($E_{acc}$) for B9AS-AES-003 cavity after the 120 °C bake, mid-T bake, ultralight EP treatments applied following mid-T baking.

## CONCLUSION

In this study, the 1.3 GHz and 650 MHz single-cell cavities, which showed high accelerating gradients in the baseline tests, were selected for mid-T baking at 350 °C for 3 h. The 1.3 GHz cavity showed a high $Q_o$ compared to the baseline. However, its *Q-E* curve lacked the typical anti-Q slope, and the quench field was significantly reduced to 22 MV/m. The 650 MHz cavity showed slightly lower $Q_o$ compared to its baseline performance. The reduced $Q_o$ in both cavities was mainly linked to an increase in $R_{res}$ caused by surface impurities, most likely NbC, as indicated by SIMS analysis of samples subjected to the same mid-T bake. The role of surface impurities in lowering $Q_o$ was confirmed when ultralight EP that removed the surface contaminants from the RF layer improved $Q_o$ in both cavities. The quench field in the 1.3 GHz cavity also increased from 22 to 32 MV/m after ultralight EP. The results demonstrate that ultralight EP is an effective post-bake treatment to recover performance losses associated with surface contaminants formed during mid-T baking.